\pdfoutput=1
\documentclass[aps,jap,twocolumn,showpacs,superscriptaddress]{revtex4}

\usepackage{amsmath}
\usepackage{graphicx}
\usepackage{gensymb}
\usepackage{float}

\begin{document}
	\title{Cation interstitial diffusion in lead telluride and cadmium telluride studied by means of neural network potential based molecular dynamics simulations}

	\author{Marcin Mi{\'n}kowski}
	\affiliation{Faculty of Physics, University of Vienna, Boltzmanngasse 5, 1090 Vienna, Austria}
	\author{Kerstin Hummer}
	\affiliation{Faculty of Physics, University of Vienna, Boltzmanngasse 5, 1090 Vienna, Austria}
	\author{Christoph Dellago}
	\affiliation{Faculty of Physics, University of Vienna, Boltzmanngasse 5, 1090 Vienna, Austria}
	\affiliation{Vienna Research Platform on Accelerating Photoreaction Discovery, University of Vienna, Währinger Str. 17, 1090 Vienna, Austria}

	\begin{abstract}
		Using a recently developed approach to represent ab initio based force fields by a neural network potential, we perform molecular dynamics simulations of lead telluride (PbTe) and cadmium telluride (CdTe) crystals. In particular, we study the diffusion of a single cation interstitial in these two systems. Our simulations indicate that the interstitials migrate via two distinct mechanisms: through hops between interstitial sites and through exchanges with lattice atoms. We extract activation energies for both of these mechanisms and show how the temperature dependence of the total diffusion coefficient deviates from Arrhenius behaviour. The accuracy of the neural network approach is estimated by comparing the results for three different independently trained potentials.
	\end{abstract}
	
	\maketitle
	
	\section{Introduction}
	Lead telluride (PbTe) and cadmium telluride (CdTe) are semiconductors that combined constitute an immiscible system, which is often employed in the manufacturing of various nanostructures such as two-dimensional nanolayers, one-dimensional nanowires and zero-dimensional quantum dots \cite{Heiss1,Koike1}. Experiments show that even at very high temperatures, but below the melting point, these two compounds remain separated and morphological transformations of the shape of the particular phases occur, both, during annealing of an as-grown sample \cite{Groiss1} and multilayer crystal growth \cite{Karczewski1}.
	
	By the appropriate choice of the experimental conditions, such as the thickness of the deposited layers, growth rate and temperature, one can design nanoobjects of desired shape and size \cite{Karczewski1,Minkowski1,Minkowski2}. From the application point of view, the most interesting structures are PbTe quantum dots and quantum wells embedded in CdTe, which can emit and detect light in the mid-infrared range \cite{Heiss1,Koike1,Bukala1}.
	
	Despite many experimental studies concerning PbTe/CdTe growth, the theoretical understanding of the dynamics of the underlying processes occurring during the morphological transformations is scarce. The multistage disintegration of a single PbTe layer inserted between two bulk CdTe materials into separate quantum dots during anneling at high temperatures was reproduced by the Cahn-Hilliard model \cite{Groiss1}. Furthermore, the growth of PbTe/CdTe multilayers was studied by means of kinetic Monte Carlo \cite{Minkowski1,Minkowski2}. However, the models used in these studies have coarse-grained character without explicitly taking into account the underlying processes on the microscopic level. Hence, the atomistic mechanism of the morphological transformations across the PbTe/CdTe interface is still unknown.
	
	On the atomic scale one can expect that the morphological transformations described above are the result of diffusion of atoms across the PbTe/CdTe interface \cite{Leitsmann1}. Point defects such as interstitials or vacancies could play a crucial role. For instance, interstitials from one compound could diffuse across the interface to the other one and participate there in the transformation of the crystal structure. Therefore, the first step in studying the mechanism of the morphological transformations is to focus on diffusion of native defects in PbTe and CdTe, without considering the interface. 
	
	Molecular dynamics (MD) simulations are a common method for studying the time evolution of condensed matter systems. In order to perform such simulations one needs to determine the forces acting on the atoms, which are often described by empirical potentials. There exist empirical potentials for PbTe \cite{Chonan1,Qiu1}, and CdTe \cite{Wang1,Oh1} crystals. However, the accuracy of empirical potentials is in general not adequate due to their limited flexibility. Alternatively, ab initio MD simulations can be performed, which are, however, computationally expensive and therefore not practical for the analysis of the diffusion of defects, where large systems and long simulation times are required. In order to overcome the difficulties related to both these approaches, we employ the neural network method of Behler and Parrinello \cite{Behler1, Behler2, Behler3}, in which a high-dimensional artificial neural network is trained with energies and forces determined by means of ab initio calculations. It has been shown that such neural networks are capable of accurately reproducing the references data and deliver the accuracy of ab initio methods at a fraction of their cost. To date, this approach has been applied to study properties of various systems \cite{Behler4,Khaliullin1,Morawietz1,Singraber1}.
	
	In this work, neural networks for bulk PbTe and CdTe are trained and used to carry out extensive molecular simulations for the diffusion of single cationic interstitials in these materials over a range of temperatures. From the obtained MD trajectories we calculate the diffusion coefficients for the interstitials as a function of temperature and analyse the different diffusion mechanisms occurring in the two systems. For both materials, the same calculations are performed for three independently trained neural network potentials in order to estimate the overall accuracy of the method.
	
	The remainder of the article is organized as follows. In Sec. \ref{Section:Systems} we review the main physical properties of PbTe and CdTe crystals. Then, in Sec. \ref{Section:Methods}, we introduce the computational techniques that we employed in our study and in Sec. \ref{Section:Results} we present the results of our simulation. Finally, in Sec. \ref{Section:Conclusions} we conclude by discussing our findings.

	\section{Systems}
	\label{Section:Systems}
	PbTe and CdTe are semiconductors belonging to the groups IV-VI and II-VI, respectively. They possess very similar lattice constants, however, their lattice structures are distinct. According to diffraction experiments the lattice constant $a$ of PbTe at room temperature is 6.46 {\AA} \cite{Bali1} and that of CdTe is 6.48 {\AA} \cite{Strauss1}. The lattice constant mismatch is therefore very small. However, there exists a lattice-type mismatch. PbTe crystallizes in the rocksalt structure and CdTe in the zinc-blende structure. Their lattices can be in fact represented as two interpenetrating face-centered cubic (fcc) cation and anion sublattices shifted with respect to each other by $a/2$ for rocksalt and $a/4$ for zinc-blende. Because of the different crystallographic structures the coordination number for PbTe and CdTe is also different. In PbTe each atom has 6 nearest neighbours, while in CdTe each atom has 4 of them.
	
	From annealing and growth experiments it is known that systems containing both PbTe and CdTe are immiscible, i.e. instead of mixing they remain separated in thermal equilibrium \cite{Groiss1,Karczewski1}. Consequently, an interface is created, with which some surface free energy is associated \cite{Leitsmann1}. The ratios of the energies for different interface orientations controls the shape of the nanoobjects that emerge in the experiments. Among these nanoobjects there are nanolayers, nanowires and quantum dots. The equilibrium shape of the PbTe quantum dots embedded in CdTe was predicted using ab initio methods \cite{Leitsmann1} and their electronic properties were also studied \cite{Leitsmann2}.
	
	Since PbTe and CdTe share the common anion (Te), the morphological transformations at the atomic scale must occur through the diffusion of cations (Pb and Cd) across the interface and the subsequent reorganization of the local lattice structure. It has been suggested that point defects play an important role in this process of cation exchange \cite{Li1}.
	
	There are several possible types of point defects in the crystal structure: vacancies, interstitials, Schottky and Frenkel defects. Vacancies and interstitials are the simplest defects, however, the barrier for the diffusion of vacancies in PbTe and CdTe is very high \cite{Roehl1}. Therefore, in this work we focus only on interstitials. We consider native interstitials diffusing in the bulk of PbTe and CdTe. Since we study PdTe and CdTe separately, our simulated systems do not contain interfaces between the two materials. 
	
	In PbTe crystals all energetic minima for Pb interstitials are equivalent. Each Pb interstitial atom is surrounded by 4 nearest-neighbour Pb and 4 nearest-neighbour Te lattice atoms \cite{Li1}. In CdTe, on the other hand, there are two non-equivalent energetic minima for Cd interstitials. One of them is a tetrahedrally coordinated site with Te atoms and octahedrally coordinated with Cd atoms, the other one has the reversed configuration of the first one \cite{Roehl1}. They are labelled $T_{a}$ and $T_{c}$, respectively.
	
	\section{Methods}
	\label{Section:Methods}
	In order to study the diffusion of cation interstitials in PbTe and CdTe, the following computational strategy is employed. First, the Vienna Ab initio Simulation Package (VASP) \cite{Vasp} is used to calculate the reference data, i.e. electronic ground state energies and forces for a given set of configurations of PbTe and CdTe with density-functional theory (DFT). These energies and forces are then used to train neural networks by means of the Neural Network Potential Package (n2p2) \cite{Singraber2}. Finally, the trained neural networks are used for performing MD simulations with LAMMPS \cite{Plimpton1}. In the post-processing of the trajectories obtained at different temperatures the position of the interstitial is identified at each time step and from the interstitial's trajectory its diffusion coefficient is extracted.
	
	This approach is used iteratively, that is the training set is repeatedly expanded with new configurations, for which new energies and forces are calculated, and the neural network is retrained at each iteration with the expanded set. These retrained neural networks are subsequently used to perform new MD simulations.
	
	Below we describe the methods used in this work in more detail.
			
	\subsection{Density-functional theory}
	
	All our DFT calculations are carried out with the exchange-correlation functional PBEsol \cite{Perdew}. The spin-orbit coupling is not considered. As the starting point for generating the set of reference configurations, 4$\times$4$\times$4 PbTe and CdTe supercells with the experimental lattice constants were used. Subsequent structures were created by increasing and decreasing the lattice constant, and randomly displacing the atoms from their positions in the perfect lattices. As the neural network was trained, configurations obtained from the MD trajectories were also subsequently added. Supercells containing one interstitial atom were also included in the training set such that the total number of atoms in a structure was either 512 or 513.

	The Brillouin zone integrations for all structures were performed using the $\Gamma$-point only. A convergence test for the $k$-point mesh revealed that the forces obtained with a denser 3$\times$3$\times$3 $k$-mesh did not significantly differ from the $\Gamma$-point calculations but lasted roughly 10 times longer.
		
	\subsection{Neural network potential}
	\label{subsection:NNP}
	The data obtained from VASP then can be used to train a neural network potential, which represents the force field used in the MD simulations. The input of the neural network is an atomic configuration and the output is its electronic ground state energy. Since the neural network is an analytic function of the atomic coordinates, the forces acting on each of the atoms can be determined by simple differentiation. 

	For training the neural network potential and employing it in MD simulations we use the approach introduced by Behler and Parrinello \cite{Behler1}. In this approach, the information about the atomic configurations is encoded in so called symmetry functions \cite{Behler5}. They transform the Cartesian coordinates into values that are invariant with respect to the translation and rotation of the whole system, and to the exchange of two arbitrary atoms of the same type. The values of these symmetry functions serve as input for the neural network. 
	
	The first thing to consider in constructing suitable symmetry functions is the cutoff radius $R_{c}$, which defines the size of the local environment around a given atom. Its value should be selected in such a way that all the relevant interactions are taken into account. Here $R_{c}=6$ \AA$ $ is chosen, which guarantees that in addition to all the nearest neighbours other atoms within a single fcc elementary cell are included in the cutoff sphere. The cutoff is implemented by cutoff functions $f(R)$, whose values go to zero beyond $R_{c}$. Several forms of the cutoff function have been proposed \cite{Singraber2}. In this work, we choose the polynomial
	\begin{equation}
	f^{\rm{poly2}}(x)=
	\begin{cases}
	x^{3}[x(15-6x)-10]+1&\text{for }x\leq 1\\
	0&\text{for }x>1
	\end{cases}
	\end{equation}
	where $x=R_{ij}/R_{c}$ and $R_{ij}$ is the distance between the pair of atoms $i$ and $j$. The derivatives of this particular function are continuous at the cutoff radius up to second order, such that forces change continuously as atoms move in and out of the cutoff sphere.

	Since the purpose of the symmetry functions is to provide a structural fingerprint of the environment of every atom in the system within a certain cutoff radius, they depend on the relative position of the given atom (called the central atom) and each of its neighbours. Symmetry functions are classified either as radial or angular \cite{Behler5}. Radial symmetry functions depend only on the distances between the central atom and its neighbours and are expressed as a sum of two-body functions. On the other hand, angular symmetry functions depend additionally on the angle spanned by the central atom with each pair of its neighbouring atoms. Therefore, they are expressed as a sum of three-body functions.
	
	It is important to note that since both our systems contain two different elements, separate symmetry functions must be provided to describe the distribution of atoms of each element and with each of them as the central atom. For radial functions there are four possibilities, whereas for angular functions there are six possibilities, i.e. two types of central atoms times two types of neighbour atoms, and two types of central atoms times three possible types of neighbour pairs, respectively.
		
	We choose the following radial symmetry function
	\begin{equation}
	G_{i}^{2}=\sum_{j}e^{-\eta R_{ij}^{2}}f_{c}(R_{ij}),\label{radial_symmetry_function}
	\end{equation}
	which is a sum of Gaussian functions multiplied by the cutoff function. The parameter $\eta$ determines the width of the Gaussian. Depending on the value of $\eta$, the function has a specific range of arguments in which it changes most steeply and is therefore most sensitive to changes of the interatomic distances. We choose the values of $\eta$ in such a way that the radial symmetry functions are equally spaced and the whole relevant range of interatomic distances is covered, $\eta=0.0001;0.016;0.4;0.07;0.12;0.2;0.3;0.5;0.9$\AA$^{-2}$.
	
	In addition to the radial functions we also choose the following angular symmetry functions
	\begin{align}
	G_{i}^{3}=&2^{1-\zeta}\sum_{j,k\neq i}(1+\lambda\cos\theta_{ijk})^{\zeta}e^{-\eta(R_{ij}^{2}+R_{ik}^{2}+R_{jk}^{2})}\nonumber\\
	&\times f_{c}(R_{ij})f_{c}(R_{ik})f_{c}(R_{jk}),\\
	G_{i}^{9}=&2^{1-\zeta}\sum_{j,k\neq i}(1+\lambda\cos\theta_{ijk})^{\zeta}e^{-\eta(R_{ij}^{2}+R_{ik}^{2})}\nonumber\\
	&\times f_{c}(R_{ij})f_{c}(R_{ik}).
	\end{align}
	Here, $\theta_{ijk}$ is the angle spanned by the atoms $i$, $j$ and $k$ with the atom $i$ at the center. The functions  $G_{i}^{3}$ and $G_{i}^{9}$ are called narrow and wide symmetry functions, respectively \cite{Singraber2,Singraber3}. In contrast to $G_{i}^{3}$, in the case of $G_{i}^{9}$ no cutoff function acts on the distance  between the atoms $j$ and $k$, such that the contributions of wider angles are not suppressed. The radial parts of the angular symmetry functions are identical with those of the radial symmetry functions. Their angular parts, on the other hand, depend on the angular distribution of the neighbouring atoms. Therefore, they are in fact sums of three-body functions. The parameter $\lambda$ is either 1 or -1, which sets the maximum of the angular part at $0\degree$ and $180\degree$, respectively. The value of $\zeta$, on the other hand, controls the width of the function. We choose angular symmetry functions with $\lambda=\pm 1$ and with $\zeta=1$, therefore with only two different angular parts. The values of $\eta$ in the radial part are chosen the same as for the radial functions (\ref{radial_symmetry_function}).
	
	The values of the symmetry functions for a given configuration are passed as the input to the neural network, which gives the total energy of that configuration as the output. Between the input and the output layer are hidden layers, which consist of nodes. The value at each node depends on the values in the nodes in the preceding layer. The particular way in which the values from one layer are transformed into the values in the nodes in the next layer is determined by the parameters of the neural network, called weights and biases and by the choice of the activation functions.
	
	A neural network architecture with two hidden layers, each of which has 25 nodes is chosen. The contribution of the atom $i$ to the total energy of the system is given by
	\begin{align}
	E_{i}=&f_{1}^{3}\left\{b_{1}^{3}+\sum_{l=1}^{N}a_{l1}^{23}\cdot f_{l}^{2}\left[b_{l}^{2}+\sum_{k=1}^{N}a_{kl}^{12}\right.\right.\nonumber\\
	&\left.\left.\cdot f_{k}^{1}\left(b_{k}^{1}+\sum_{j=1}^{j_{\rm max}}a_{jk}^{01}\cdot G_{j}^{i}\right) \right]\right\}.
	\end{align}
	Here $G_{j}^{i}$ is the $j$th symmetry function centered at the atom $i$, $j_{max}$ is the total number of symmetry functions for each atom, $a_{ij}^{kk+1}$ is the weight of the contribution of node $i$ in layer $k$ to node $j$ in layer $k+1$, $b_{i}^{j}$ is the bias for node $i$ in layer $j$ and $f_{i}^{j}$ is the activation function for node $i$ in layer $j$. For the first and the second layer we choose the hyperbolic tangent as activation function and for the output layer the linear function is used.
		
	The sum of energy contributions from each atom in the system $E=\sum_{i}E_{i}$ is the total potential energy of the given configuration. The force $\mathbf{F}_i$ acting on atom $i$ is obtained by calculating the gradient of the total potential energy with respect to the coordinates of atom $i$, $\mathbf{F}_i=-\nabla E$. For a three-dimensional system with $N$ atoms, there are in total $3N$ force components (3 for each atom). 
	
	During the training of the neural network, the weights and biases are adjusted to minimize the difference between the energies and forces predicted by the neural network and the reference data. This difference is quantified by the cost function
	\begin{equation}
	\Gamma=\sum_{i=1}^{n}(E_{i}^{\rm ref}-E_{i}^{\rm NN})^{2}+\beta^{2}\sum_{i=1}^{n}\sum_{j=1}^{N_{n}}(\mathbf{F}_{ij}^{\rm ref}-\mathbf{F}_{ij}^{\rm NN})^{2},\label{cost_function}
	\end{equation}
	where indices $i$ and $j$ enumerate configurations and atoms, respectively. The total number of configurations is $n$, each of which contains $N_{n}$ atoms. The parameter $\beta$ tunes the importance of the forces with respect to the energies. For the purpose of training the reference data are divided into a training set and a test set. This is done to ensure that no overfitting occurs, that is the error of the test set should be comparable with that of the training set. The neural network training proceeds in 50 epochs, during each of which all the parameters of the neural network are updated. Since the number of forces exceeds by far the number of energies, we use only a 0.0008 fraction of all the forces for the training so the number of energy and force updates is similar. That fraction is randomly selected at the beginning of each epoch.

	For optimization of the cost function (\ref{cost_function}) we use the extended Kalman filter \cite{Kalman1,Kalman2,Smith1}, which has been implemented in n2p2 \cite{Singraber3}. The Kalman filter is an algorithm originally used for estimating a dynamical system's unknown state based on a series of noisy measurements. Later it was shown that it can be also used for training neural networks \cite{Singhal1}.

	\subsection{Molecular dynamics}
	
	Once the neural network for a given system is trained, it can be used for performing MD simulations. The program n2p2 provides an interface to LAMMPS \cite{Singraber2}, which is used for all MD simulations in the present work.
	
	Since our aim is to determine the diffusion coefficient of a cation interstitial, we carry out MD simulations for 4$\times$4$\times$4 PbTe and CdTe supercells with an additional atom inserted in the middle of a unit cell. In total, the systems consisted of 513 atoms. The simulations are performed with a timestep of 2 fs with a Nose-Hoover thermostat and barostat for temperatures ranging from 700 to 1200 K in intervals of 50 K. At each run the system is first equilibrated for100 ps. Then the simulation is continued for several nanoseconds and configurations are stored every 200 fs. Both for PbTe and CdTe, the MD runs lasted 8 ns for temperatures up to 800 K, 6 ns for 850 K and 4 ns for all the higher temperatures. 

	Due to an indirect mechanism of diffusion, in which the interstitial kicks out a lattice atom that subsequently becomes the new interstitial, it is important to identify the interstitial atom at each step. This is done by defining the interstitial as the atom with the largest number of neighbours. At high temperatures, this method may be not very reliable due to large fluctuations. Therefore, the event of the exchange of the interstitial atom in only considered in case it remains for ten consecutive steps. Otherwise the exchange is considered as transient fluctuation and ignored.
	
	Using this criterion to identify the interstitial, the trajectory of the interstitial is determined and the the periodic boundary conditions are unwrapped.  After $n$ steps, the mean square displacement (MSD) \cite{Qian1,Michalet1} is estimated using 
\begin{equation}
	\langle\mathbf{r}^2(n)\rangle=\frac{\sum_{i=1}^{N-n}(\mathbf{r}_{i+n}-\mathbf{r}_{i})^2}{N-n}.\label{MSD}
\end{equation}
Here, $\mathbf{r}_i$ is the position of the interstitial after $i$ steps with respect to its starting position and $N$ is the total number of collected configurations. Expression (\ref{MSD}) takes into account that for a time lag of $n$ steps, there are $N-n$ positions that can be used to determine the MDS. The estimated MSD is more accurate for shorter times, since there are more short time lags than long ones.

	In order to extract the diffusion coefficient from the MSD, expression (\ref{MSD}) is plotted as a function of time $t$. By fitting a linear function to these data, the diffusion coefficient $D$ is obtained from the slope according to 
\begin{equation}
	D=\frac{\text{MSD slope}}{2d},
\end{equation}
where $d$ is the number of dimensions. In our case $d=3$.
	
	With this procedure the diffusion coefficient can be studied in  the  whole  range  of  temperatures. This further allows to calculate the activation energy by fitting the data to the Arrhenius law.  Moreover, the number of jumps in the trajectory is counted by analysing the displacements of the interstitial. In doing so, two different types of diffusion jumps are distinguished, i.e., exchange and direct jump.

	\section{Results}
	\label{Section:Results}
	\subsection{Neural network training}
	
	For each of the two investigated materials, PbTe and CdTe, three independent neural networks with different random seeds chosen for the initialization of the values of the network’s parameters were trained. Consequently, the final values of these parameters were different for each network also at the end of the training, even though all three networks represented the same system.
	
	The configurations used for the training were created iteratively. First, the lattice constants of the PbTe rocksalt and CdTe zinc blende unit cells were optimized using VASP. The initial set consisted of a few hundreds of 4x4x4 supercells created with these optimized lattice constants. In addition to the perfect structures, configurations with the atoms displaced randomly from their equilibrium positions were also used. To the subsequent sets configurations with slightly modified lattice constants were added to account for the thermal expansion of crystals at finite temperatures. Moreover, new configurations were also taken from MD trajectories. During the MD simulations, an extrapolation warning was produced each time the value of some symmetry function was out of range for which the neural network was initially trained. The number of such warnings was monitored during the simulations and the configurations with the largest number of extrapolation warnings were added to the set. Additionally, predictions for the same trajectory by two independently trained neural networks were compared and those configurations for which the predictions differed considerably were also added. This procedure was repeated until there were no extrapolations warnings and no significant differences between different neural networks in the relevant range of temperatures.

	After training the neural networks for defect-free crystals, a single cation interstitial was introduced to the systems, that is Pb in PbTe and Cd in CdTe. This required further ab initio calculations and further retraining of the neural network. For creating configurations with an interstitial the same procedure was used as for the defect-free crystal. Special attention was paid to configurations with the interstitial around the transition states. They were identified visually and added to the training set. During the MD simulations there are fewer such configurations than those with the interstitial around one of the energetic minima but the former are very important for reproducing the correct energy barrier for the diffusion.
	
	In total 4898 configurations of PbTe and 2866 of CdTe were generated, of which  90\% were used as training set and 10\% as test set. For the training all the energies from the test set were used and 0.08 \% of the forces, which gave a comparable number of energy and force updates in each epoch. The relative importance of the forces was set to $\beta=5$. The training proceeded for 50 epochs and the weights and biases from the last epoch were used for the MD simulations.
	
	\begin{table}[h!]
		\begin{tabular}{c|c|c}
			\toprule
			\textbf{NNP} & \textbf{energy RMSE} & \textbf{force RMSE}\\
			&[meV/atom]&[meV/\AA]\\
			\hline
			PbTe nnp-1 & 0.493/0.569 & 72.7/68.9\\
			PbTe nnp-2 & 0.475/0.568 & 72.9/71.3\\
			PbTe nnp-3 & 0.498/0.569 & 70.5/70.3\\
			CdTe nnp-1 & 0.255/0.381 & 54.8/63.3\\
			CdTe nnp-2 & 0.254/0.365 & 59.7/61.6\\
			CdTe nnp-3 & 0.250/0.544 & 56.9/72.3
		\end{tabular}
		\caption{Root mean square errors for energies and forces of the neural networks trained in this work. The first value in each of the pairs is the error of the training set and the second one is that of the test set.}
		\label{RMSE}
	\end{table}

	In Table \ref{RMSE} the root mean square errors (RMSE) for all the neural networks trained and used subsequently in this work are listed. They are comparable to RMSE obtained previously with neural network potentials for other materials \cite{Singraber3,Morawietz1}. In all cases the RMSE is similar and the error for the training and the test set is of the same order. We notice that for CdTe the errors are always smaller than for PbTe.
	
\subsection{Diffusion of cation interstitials}
	
	\begin{figure}[b]
		\centering
		\hspace*{-0.5cm}
		\includegraphics[width=10cm]{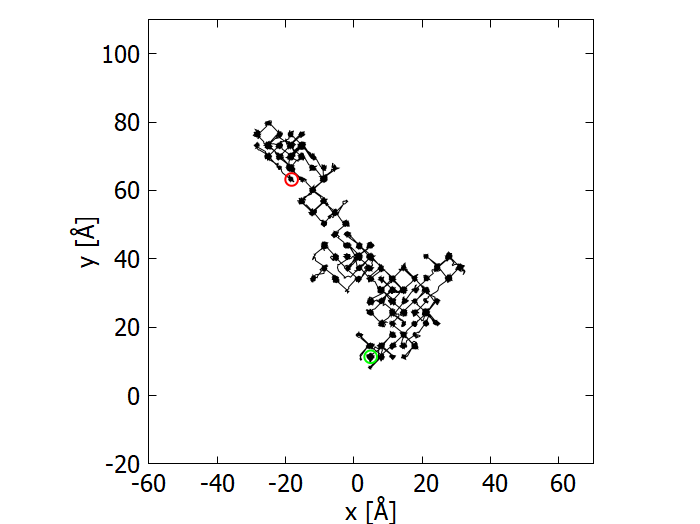}
		\hspace*{-0.5cm}
		\includegraphics[width=10cm]{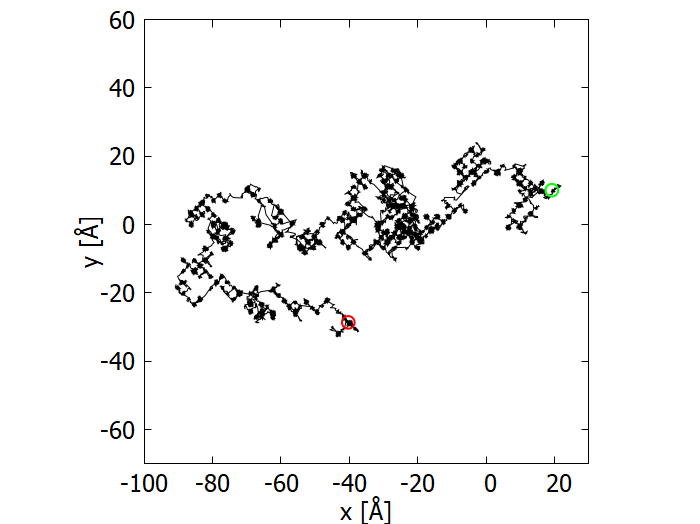}
		\caption{Trajectory of a Pb interstitial in PbTe (top) and a Cd interstitial in CdTe (bottom) at $T=700$K for nnp-1 projected on the $xy$-plane. The green and red circles denote the beginning and the end of the trajectory, respectively.}
		\label{trajectories}
	\end{figure}

	\begin{figure}
		\centering
		\includegraphics[width=9cm]{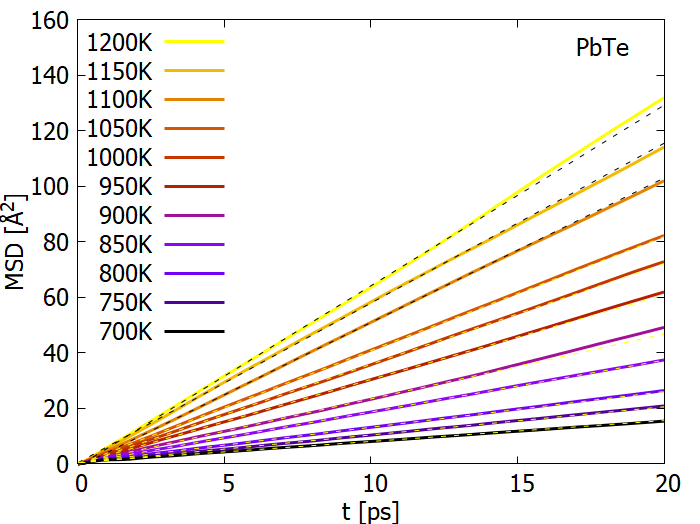}
		\includegraphics[width=9cm]{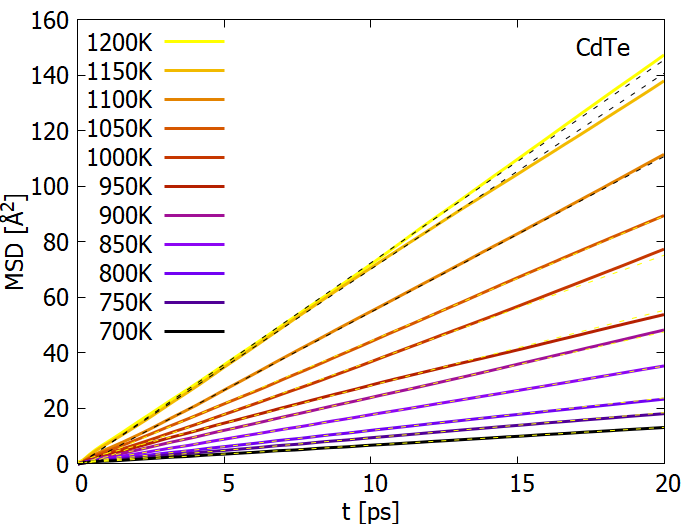}
		\caption{Mean squared displacement measured for a single Pb interstitial in PbTe (top) and a single Cd interstitial in CdTe (bottom) at temperatures in the range from 700 K to 1200 K for nnp-1. The dashed lines represent a linear fit to the MSD curve in the range 2-14ps, from which the diffusion coefficient has been extracted.}
. 
		\label{MSD_plot}
	\end{figure}
	\begin{figure}
		\centering
		\hspace*{-0.5cm}
		\includegraphics[width=9cm]{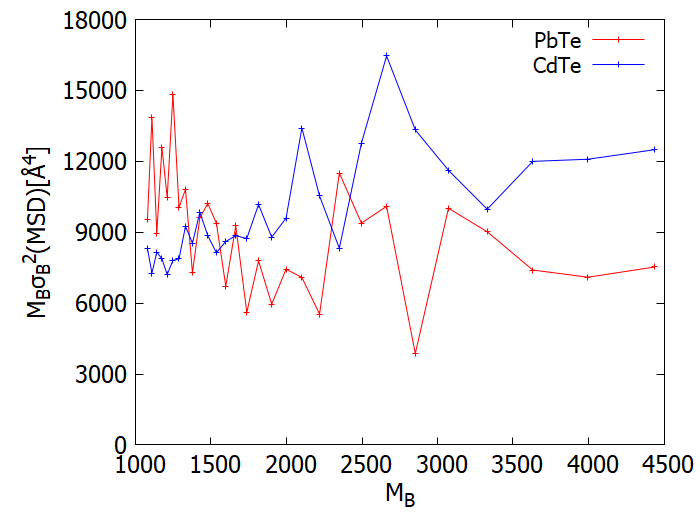}
		\caption{Block averaging for the MSD at 10 ps of a single interstitial at 700 K in PbTe (red) and CdTe (blue) for nnp-1. The product of the block size and the variance of the average MSD reaches a plateau when plotted as a function of the block size.}
		\label{block_average}
	\end{figure}
	
	\begin{figure}[b]
		\centering
		\includegraphics[width=9cm]{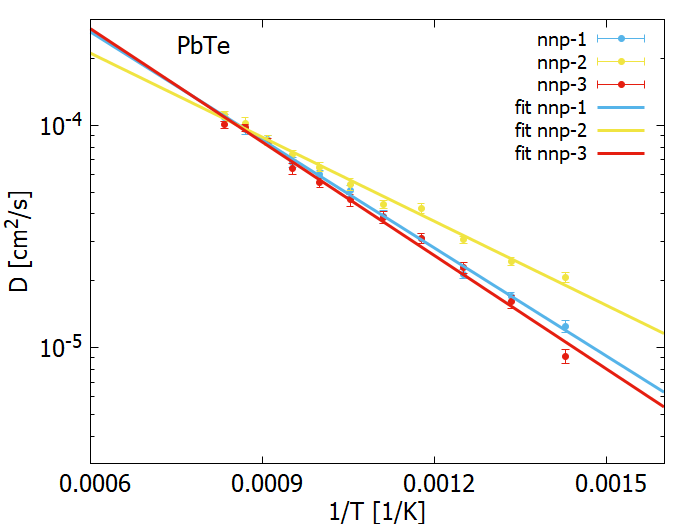}
		\includegraphics[width=9cm]{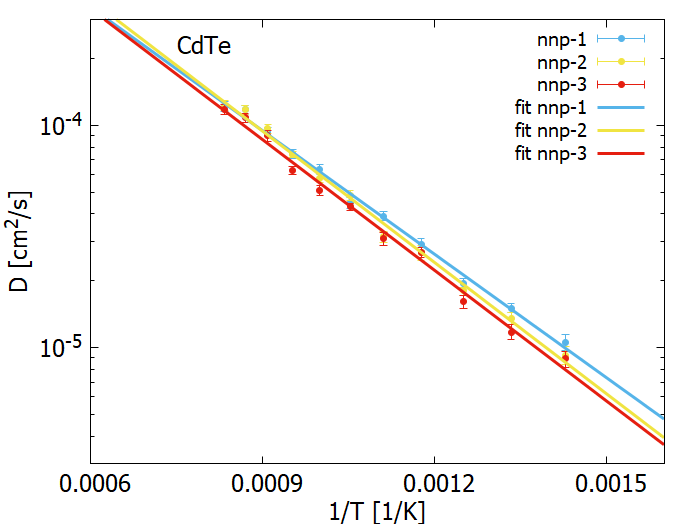}
		\caption{Diffusion coefficient for a Pb interstitial in PbTe (top) and a Cd interstitial in CdTe (bottom) for three different neural network potentials fitted to the Arrhenius law.}
		\label{linear_fit}
	\end{figure}

	During the MD simulations the interstitial atom diffuses in the supercell. No creation of other defects due to thermal fluactions was observed, not even at the highest temperature studied, 1200 K. Examples of trajectories of the defect obtained from the MD simulations are shown in Fig. \ref{trajectories}. Along these trajectories, two distinct mechanisms of diffusion were observed. One of them involved simple hops of the diffusing interstitial atom between neighbouring interstitial sites, while the other mechanism was an exchange of the interstitial atom with one of the lattice atoms. During the latter process the interstitial took over the position of the lattice atom, which subsequently became a new interstitial. In PbTe and CdTe, both, hops and exchanges were observed. However, in PbTe exchanges represent the dominant mechanism, whereas in CdTe hops are the preferred one. Both mechanisms will be discussed in more detail in the next subsection.
	
	In Fig. \ref{MSD_plot} the mean square displacement for a single cation interstitial in PbTe and CdTe is shown as the function of the time lag. The plots correspond to the trajectories generated at various temperatures for the potential nnp-1. As expected, for short time lags the dependence of MSD on time is linear. For longer times the averaging is done over fewer time lags with more correlations leading to larger statistical errors. Therefore, in order to obtain the diffusion coefficient we fitted a linear function to the linear part of the MSD curve, that is for short times between 2 and 14 ps. The same procedure was applied to extract the diffusion coefficient in all the trajectories studied.
	
	Moreover, the convergence of the method applied is tested by means of block averaging \cite{Newman1,Rapaport1}. For this purpose, the series of instantaneous square displacements at 10 ps, for which the MSD is still linear, is obtained for the whole trajectory. This series corresponds to the subsequent terms in the sum in Eq. (\ref{MSD}) and contains $M=N-n$ data points, which is equal to the number of the time lags. Then the series is divided into $n_{B}$ equal blocks, each of which has $M_{B}=M/n_{B}$ data points. For each of the blocks the MSD is extracted by taking the average over $M_{B}$ data points, yielding $MSD^{(i)}_{B}$. Then the average over all block averages is calculated as
	\begin{equation}
		\langle MSD\rangle_{B}=\frac{1}{n_{B}}\sum_{i=1}^{n_{B}}MSD_{B}^{(i)}.
	\end{equation}
The variance of the block average is estimated by
	\begin{equation}
		\sigma_{B}^{2}(MSD)=\frac{1}{n_{B}}\sum_{i=1}^{n_{B}}(MSD_{B}^{(i)}-\langle MSD\rangle_{B})^{2}.
	\end{equation}
If the simulation is properly converged, the product $M_{B}\sigma_{B}^{2}(MSD)$ should reach a plateau $s$ in the limit $M_{B}\rightarrow\infty$. The variance of the average MSD can be then estimated as
	\begin{equation}
		\sigma^{2}(\langle MSD\rangle)=\frac{s}{M}
	\end{equation}
and the error of the diffusion coefficient is equal to
	\begin{equation}
		\Delta D=\sigma(\langle MSD\rangle)/{6t},
	\end{equation}
where $t=10$ ps.

In Fig. \ref{block_average} the method is illustrated for nnp-1 for the instersitial trajectories in PbTe and CdTe at the temperature of 700 K. The product of the block size and the variance of the average MSD is plotted as the function of the block size. For both systems it reaches a plateau around $M_{B}=3500$. It corresponds to the values of $s$ around $8.0\cdot 10^{4}$ \AA$^{4}$ in PbTe and $1.2\cdot 10^{4}$ \AA$^{4}$ in CdTe, and the errors of the diffusion coefficient $7.458\cdot 10^{-7}$ cm$^{2}$/s and $9.134\cdot 10^{-7}$ cm$^{2}$/s, respectively. The same procedure is performed in the whole range of the temperatures studied and for all three neural network potentials.

	\begin{figure*}
		\centering
		\includegraphics[width=18cm]{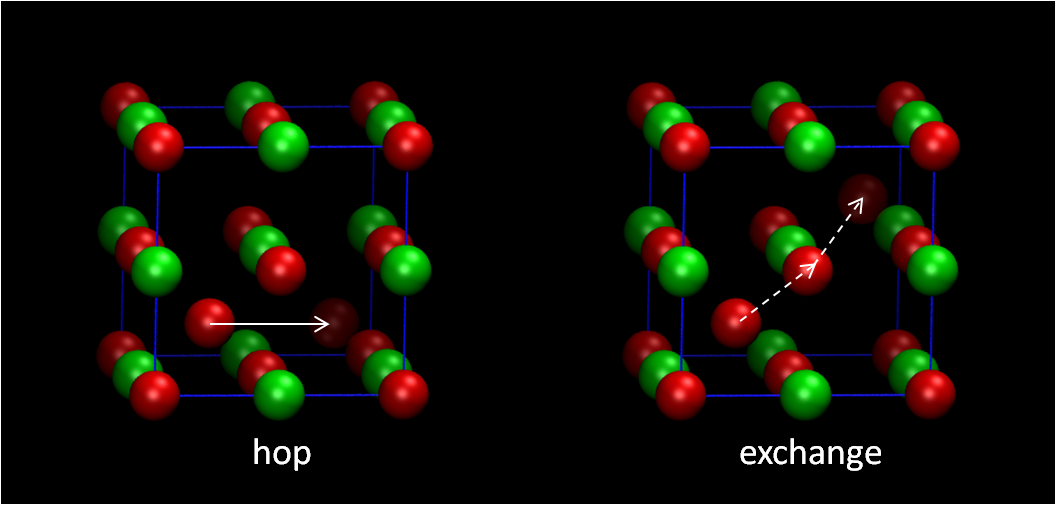}
		\caption{Scheme of the hop and the exchange mechanism of diffusion of an interstitial Pb atom in PbTe. The arrows represent the jumps of the atoms involved in the respective mechanism. The final position of the interstitial atom is shown with the transparent colour. During a hop the interstitial Pb atom jumps along one of the [100] directions between energetically equivalent minima. On the other hand, during an exchange the interstitial Pb atom replaces one of the nearest lattice Pb atoms, which subsequently becomes the new interstitial. As seen in the scheme, one exchange can be equivalent to two or three hops.}
		\label{PbTe_mechanisms}
	\end{figure*}
	
	\begin{figure*}
		\centering
		\includegraphics[width=18cm]{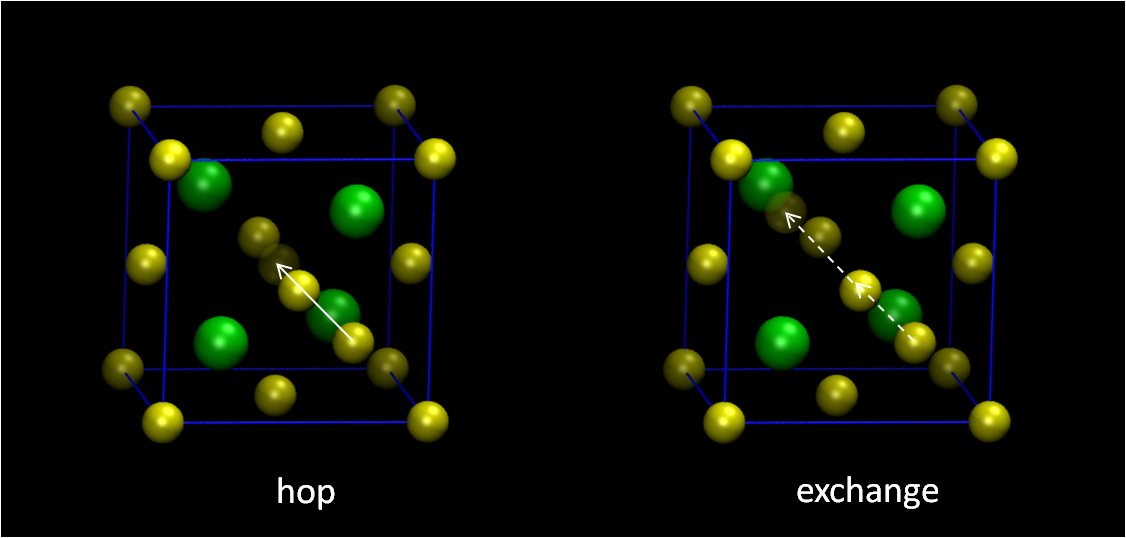}
		\caption{Scheme of the hop and the exchange mechanism of diffusion of an interstitial Cd atom in CdTe. The arrows represent the jumps of the atoms involved in the respective mechanism. The final position of the interstitial atom is shown with the transparent colour. During a hop the interstitial Cd atom jumps between $T_{c}$ and $T_{a}$ positions. On the other hand, during an exchange the interstitial replaces one of the nearest lattice Cd atoms, which subsequently becomes the new interstitial. As seen in the scheme, one exchange can be equivalent to two hops.}
		\label{CdTe_mechanisms}
	\end{figure*}

	The temperature dependence of the diffusion coefficients is shown for all neural network potentials in the Arrhenius plots in Fig. \ref{linear_fit}. The values of the diffusion coefficient at particular temperatures are represented by the dots with the corresponding error bars. Additionally, the data points are fitted with the Arrhenius equation
	\begin{equation}
	D(T)=D_{0}\exp(-\beta E_{a}),\label{activation_energy}
	\end{equation}
	where $\beta=1/(k_{B}T)$. Here, $E_{a}$ is the activation energy for the diffusion and $D_{0}$ is the diffusion prefactor. Both parameters can be extracted from the fit. The error of the activation energy is estimated by converting Eq. (\ref{activation_energy}) into
	\begin{equation}
	E_{a}=-\frac{\ln(D/D_{0})}{\beta}
	\end{equation}
and performing error propagation with respect to the diffusion coefficient
	\begin{equation}
	\Delta E_{a}=\frac{\Delta D}{\beta D}.
	\end{equation}
The above result depends on the temperature, here for the estimation of the error T=1000 K is chosen, which is the value in the middle of the range of the studied temperatures.
	\begin{table}[h!]
		\begin{tabular}{c|c|c}
			\toprule
			\textbf{NNP} & $E_a$ [meV] & $D_{0}$ [cm$^2$/s] \\
			\hline
			PbTe nnp-1 & 322 $\pm$ 4 & 2.475$\cdot$10$^{-3}$\\
			PbTe nnp-2 & 250 $\pm$ 4& 1.208$\cdot$10$^{-3}$\\
			PbTe nnp-3 & 338 $\pm$ 4& 2.862$\cdot$10$^{-3}$\\
			CdTe nnp-1 & 368 $\pm$ 5& 4.413$\cdot$10$^{-3}$\\
			CdTe nnp-2 & 391 $\pm$ 4& 5.604$\cdot$10$^{-3}$\\
			CdTe nnp-3 & 389 $\pm$ 4& 5.032$\cdot$10$^{-3}$
		\end{tabular}
		\caption{Activation energies $E_{a}$ and diffusion prefactors $D_{0}$ for Pb and Cd interstitials in PbTe and CdTe, respectively, extracted from the Arrhenius plot for each of the neural network potentials.}
		\label{diffusion parameters}
	\end{table}
	
	In Table \ref{diffusion parameters} the activation energies and the diffusion prefactors for all the Arrhenius plots in Fig. \ref{linear_fit} are collected. For PbTe the activation energies tend to be lower than for CdTe, which suggests that diffusion of interstitials in PbTe occurs more easily than in CdTe. On the other hand, the diffusion prefactors in CdTe are higher than in PbTe. Due to competition between these two effects, the diffusion coefficient is larger in PbTe at low temperatures, but at high temperatures it is higher in CdTe. For PbTe nnp-2 the diffusion prefactor is twice as low as for the other two PbTe neural network potentials. However, it is compensated by the lower activation energy for nnp-2. As can be seen in Fig. \ref{linear_fit}, the diffusion coefficient for PbTe nnp-2 at low temperature is slightly higher than for nnp-1 and nnp-3.

	\begin{figure*}
		\centering
		\includegraphics[width=18cm]{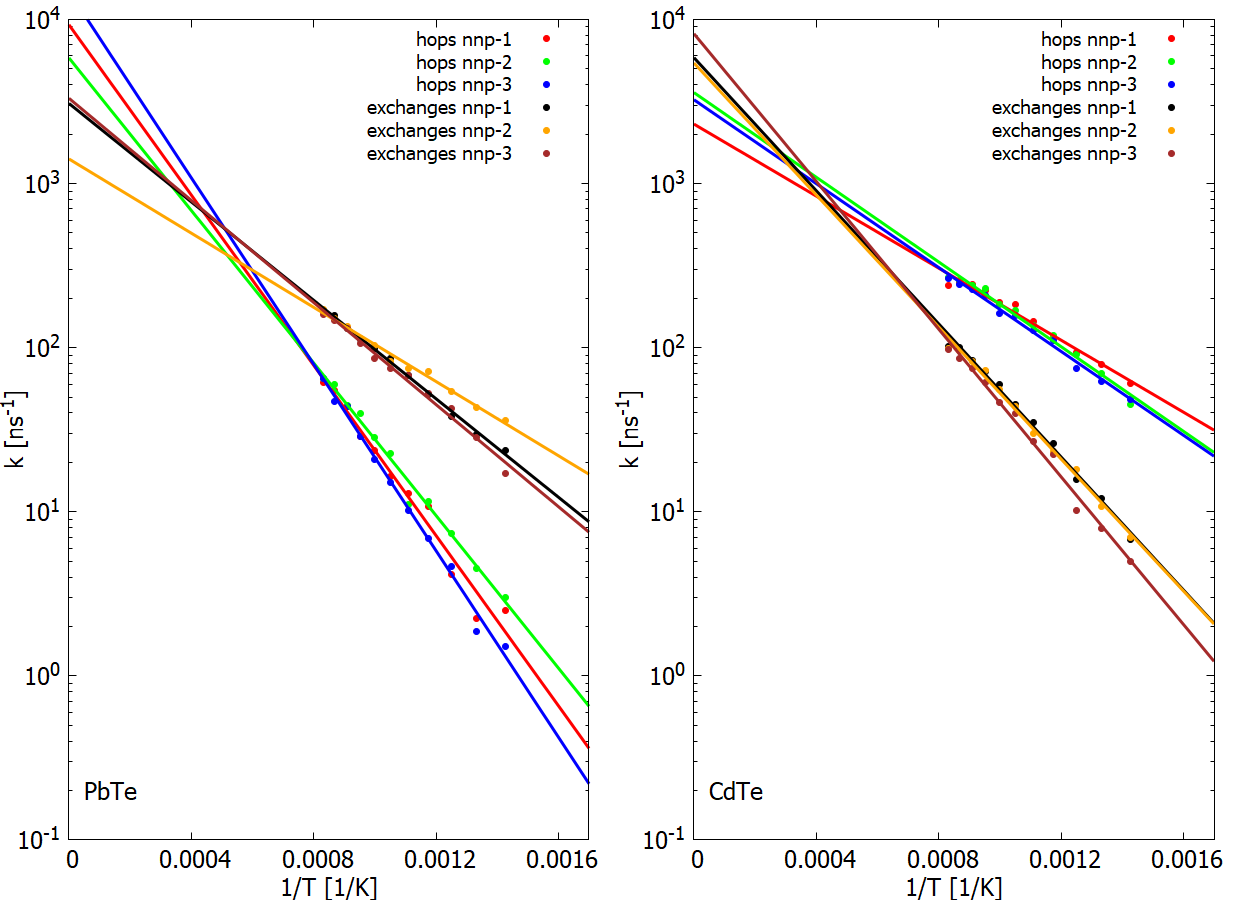}
		\caption{Arrhenius plots of hop and exchange rates in PbTe (left) and CdTe (right). The dots represent the values of these rates extracted from the MD trajectories at given temperatures. The lines are the corresponding Arrhenius fits.}
		\label{diffusion_mechanisms}
	\end{figure*}
	
\subsection{Diffusion mechanisms}
		
	As already mentioned, two different mechanisms of interstitial diffusion were observed in the MD simulations of PbTe and CdTe. In one of them the interstitial atom simply jumps between two different energetic minima, in the other one it exchanges its position with one of the lattice atoms. We refer to these mechanisms as hops and exchanges, respectively.

	The mechanisms of diffusion in PbTe are illustrated in Fig. \ref{PbTe_mechanisms}. Since there is only one equilibrium position for the interstitial atom in PbTe and two in CdTe, the diffusion in the former is simpler. During a hop, the interstitial Pb atom jumps between the nearest-neighbouring minima along the [100] direction. The length of the jump is equal to $a/2$. On the other hand, during the exchange process the interstitial Pb atom kicks out one of its four neighbouring Pb lattice atoms, which in turn assumes the position in one of the neighbouring minima, either in the [110] or [111] direction relative to the original position of the first interstitial. The effective jump of the interstitial Pb atom is therefore $(a/2,a/2,0)$ or $(a/2,a/2,a/2)$ and analogously in all the equivalent crystallographic directions. Hence, during an exchange the interstitial moves by a distance of $a\sqrt{2}/2$ or $a\sqrt{3}/2$, respectively. 
		
	Analogously, in Fig. \ref{CdTe_mechanisms} the diffusion mechanisms in CdTe are shown. The diffusion of interstitial Cd atoms is more complex. As mentioned in Sec. \ref{Section:Systems}, for the interstitial Cd atom there are two different equilibrium positions labelled $T_{a}$ (tetrahedral anion site) and $T_{c}$ (tetrahedral cation site). Moreover, the Cd interstitial can exist in three charge states  \cite{Ma1, Yang1}: one neutral and two charged ones, Cd$^{+}$ and Cd$^{2+}$. The height of the energy barrier between the interstitial sites depends on the particular charge state. In the present case the interstitial atom occupies most of the time the $T_{a}$ site and only for a very short time it is found in $T_{c}$ position, which corresponds to the state Cd$^{2+}$ discussed in Refs. \onlinecite{Ma1} and \onlinecite{Yang1}. The diffusion through hops proceeds between $T_{a}$ and $T_{c}$ sites along the [110] direction, as described in Ref. \onlinecite{Ma1}. It consists of jumps of the interstitial Cd atom between these sites in the directions [111] and [11$\overset{-}{1}$]. The length of a single jump is $\frac{\sqrt{3}}{2}a$. Exchanges, on the other hand, occur mostly between two $T_{c}$ sites, also in the [110] direction. One exchange is therefore effectively equivalent to two successive hops.

	As can be seen from the above analysis of the diffusion mechanisms, in both materials, PbTe and CdTe, the interstitial jump is longer for an exchange than for a hop. As a consequence, even though there are fewer exchanges than hops in CdTe, the contribution of both mechanisms to the total diffusion coefficient is comparable. From the MD trajectories the number of jump events for both hops and exchanges was extracted. Since the identity of the interstitial (specified in LAMMPS by its index) and its position are known at each step, a jump is counted when the particle is displaced by the length comparable with the distance between the neighbouring interstitial sites. If the index of the particle changes after the jump, it is considered as an exchange, otherwise it is a hop. The corresponding jump rates $k$ are shown as an Arrhenius plot for all considered neural network potentials and compared for PbTe and CdTe in Fig. \ref{diffusion_mechanisms}. This quantitative analysis confirms the observation that exchanges dominate in PbTe and hops in CdTe. However, for both materials their ratio becomes closer to 1 as the temperature increases. At 700 K one of the jump rates is always around one order of magnitude higher than the other one, while at 1200 K they are of the same order.

	\begin{figure}
		\centering
		\includegraphics[width=9cm]{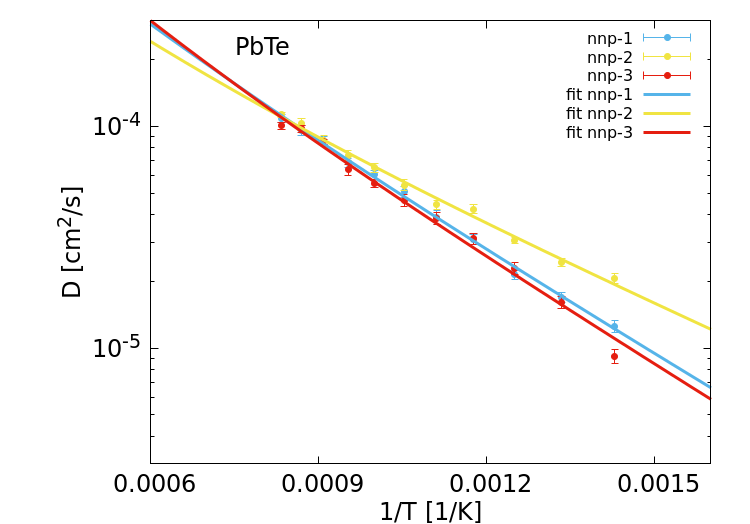}
		\includegraphics[width=9cm]{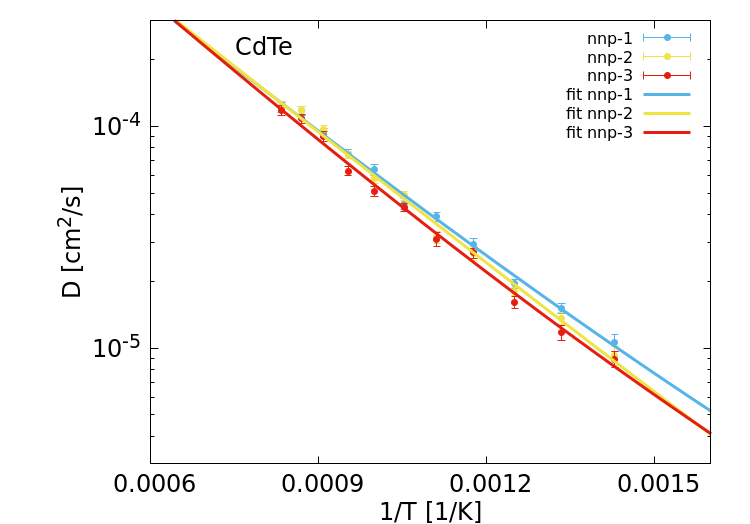}
		\caption{Diffusion coefficient for a Pb interstitial in PbTe (top) and a Cd interstitial in CdTe (bottom) for three different neural network potentials fitted to the relation (\ref{diff_coeff_curve}).}
		\label{curve_fit}
	\end{figure}
			
	\begin{table}[h]
		\begin{tabular}{c|c|c}
			\toprule
			\textbf{NNP} & E$_{\rm hops}$ [meV] & E$_{\rm ex}$ [meV] \\
			\hline
			PbTe nnp-1 & 515 & 298\\
			PbTe nnp-2 & 461 & 224\\
			PbTe nnp-3 & 564 & 309\\
			CdTe nnp-1 & 218 & 402\\
			CdTe nnp-2 & 256 & 399\\
			CdTe nnp-3 & 254 & 447
		\end{tabular}
		\caption{Activation energies for the two different diffusion mechanisms in PbTe and CdTe, hops ($E_{\rm hops}$) and exchanges ($E_{\rm ex}$), extracted from the Arrhenius plots for the corresponding jump rates.}
		\label{activation_energies_mechanisms}
	\end{table}

As can be inferred from Fig. \ref{diffusion_mechanisms}, for both mechanisms the jump rates follow the Arrhenius law. Therefore, in analogy to the total diffusion activation energy, the particular activation energies for hops and exchanges can be determined by fitting a line to the Arrhenius plots in Fig. \ref{diffusion_mechanisms}. In Table \ref{activation_energies_mechanisms}, these activation energies, $E_{\rm hops}$ for hops and $E_{\rm ex}$ for exchanges, are summarized for PbTe and CdTe. As expected, for PbTe the activation energy is higher for hops than for exchanges and for CdTe it is the opposite. For each neural network potential the effective activation energy given in Table \ref{diffusion parameters} is located between the corresponding energies listed in Table \ref{activation_energies_mechanisms}. These separate activation energies can be further used to fit a more complex function to the data than the Arrhenius law. The diffusion coefficient for a particle which jumps by means of two independent energetically activated mechanisms can be written as the sum
	\begin{equation}
	D=D_{0}^{\rm hops}\exp(-\beta E_{\rm hops})+D_{0}^{\rm ex}\exp(-\beta E_{\rm ex}),\label{diff_coeff_curve}
	\end{equation}
where $D_{0}^{\rm hops}$ and $D_{0}^{\rm ex}$ are the diffusion prefactors for hops and exchanges, respectively. The results of this fitting are summarized in Table \ref{diffusion_prefactors_mechanisms} and the corresponding curves are shown in Fig. \ref{curve_fit}. By comparing the curves with those in Fig. \ref{linear_fit}, it can been seen that considering separate mechanisms of diffusion allows for a more accurate fitting of the temperature dependence of the diffusion coefficient to the simulation data.
	\begin{table}[h!]
		\begin{tabular}{c|c|c}
			\toprule
			\textbf{NNP} & D$_{0}^{\rm hops}$ [cm$^2$/s] & D$_{0}^{\rm ex}$ [cm$^2$/s]\\
			\hline
			PbTe nnp-1 & 3.106$\cdot10^{-3}$ & 1.614$\cdot10^{-3}$\\
			PbTe nnp-2 & 2.074$\cdot10^{-3}$ & 7.502$\cdot10^{-4}$\\
			PbTe nnp-3 & 4.731$\cdot10^{-3}$ & 1.775$\cdot10^{-3}$\\
			CdTe nnp-1 & 1.146$\cdot10^{-4}$ & 5.539$\cdot10^{-3}$\\
			CdTe nnp-2 & 6.373$\cdot10^{-5}$ & 5.779$\cdot10^{-3}$\\
			CdTe nnp-3 & 2.542$\cdot10^{-4}$ & 7.331$\cdot10^{-3}$
		\end{tabular}
		\caption{Diffusion prefactors for hops ($D_{0}^{\rm hops}$) and exchanges ($D_{\rm 0}^{ex}$) of interstitials in PbTe and CdTe for the different neural network potentials.}
		\label{diffusion_prefactors_mechanisms}
	\end{table}

	Diffusion of Pb in PbTe was studied experimentally in Ref. \onlinecite{Gomez1}. The activation energy was measured to be 249 meV, which is close to the values reported in this work (322, 250, 338 meV). In contrast, the diffusion prefactor was found to be $3.1\times10^{-6}$ cm$^{2}$/s, which is three orders of magnitude smaller than our result. However, the method used in Ref. \onlinecite{Gomez1} is based on radioactive isotopes, which does not take into account the exchange mechanism.
	
	The diffusion coefficient for a Cd interstitial in CdTe was measured in Ref. \onlinecite{Wolf1}. The value obtained there for the temperature 800 K was 1.75$\cdot$10$^{-6}$ cm$^{2}$/s, which is one order of magnitude smaller than the value calculated in this work (1.95$\cdot$10$^{-5}$; 1.80$\cdot$10$^{-5}$; 1.61$\cdot$10$^{-5}$ cm$^{2}$/s).

	Moreover, diffusion of a cation interstitial atom in CdTe was studied in Ref. \onlinecite{Roehl1} by means of nudged elastic band method (NEB) \cite{Jonsson1}. The corresponding energy barrier determined there is 330 meV, which is close to the activation barriers reported in this work (368, 391, 389 meV). However, the difference between the NEB method and the approach used here is that in the former one finds the minimum energy path between two fixed endpoints, which allows to get the energy barrier but does not take into account finite temperature effects. In MD simulations the system evolves according to the equations of motion at specified external conditions (such as temperature). Moreover, because in NEB the initial and the final configurations are fixied, it is not possible to find any new diffusion mechanisms, as it was done in this work.

\section{Conclusions}
\label{Section:Conclusions}
	
	In this work, the diffusion processes of interstitial Pb and Cd atoms have been studied in a supercell of PbTe and CdTe, respectively. Futhermore, a procedure of extracting the value of the diffusion coefficient from the trajectories generated by neural network based MD simulations has been demonstrated. For the training of the neural network potentials {\em ab initio} data calculated with the PBEsol functional were used. For both systems the results for the diffusion coefficients for three different independently trained neural network potentials were compared. The corresponding activation energies were extracted and it was found that they differ from each other by no more than 100 meV, which can be viewed as the accuracy of the neural network approach used in this work for the calculation of activation energies.
	
	Both in PbTe and CdTe two different mechanisms of interstitial diffusion have been observed, namely hops and exchanges, for which separate activation energies have been extracted. Hops were more frequent in CdTe and exchanges in PbTe. However, in both materials exchanges had longer effective jump lengths. Therefore, interstitial diffusion is dominantly controlled by exchanges in both materials. Taking into account two diffusion mechanisms with different activation energies explains the deviations of the temperature dependence of the diffusion coefficient from the Arrhenius law.
	
	The mechanism of atom exchange is particularly interesting in the context of the morphological transformations observed experimentally in PbTe/CdTe systems. In this work, only self-diffusion of cations within either PbTe or CdTe has been considered. However, the framework presented here can be also used to study the diffusion of foreign atoms in the crystal. One can expect that Pb cations introduced in CdTe exchange with the host Cd cations and Cd cations in PbTe exchange with the host Pb cations. This could lead to a subsequent rebuild of the local crystal structure, which in turn could be an underlying mechanism for the morphological transformations observed in the PbTe/CdTe growth and annealing experiments. Further studies in this direction will be necessary to clarify this question.
	
	\section{Acknowledgements}
	
	The research has been supported by the Austrian Science Fund (FWF) project M 2382-N28. Calculations have been performed on the Vienna Scientific Cluster (VSC).


\begin{thebibliography}{99}
	\bibitem{Heiss1} W. Heiss, H. Groiss, E. Kaufmann, G. Hesser, M. B{\"o}berl, G. Springholz, F. Sch{\"a}ffler, Appl. Phys. Lett. \textbf{88}, 192109 (2006)
	\bibitem{Koike1} K. Koike, H. Harada, T. Itakura, M. Yano, W. Heiss, H. Groiss, E. Kaufmann, G. Hesser, F. Sch{\"a}ffler, J. Cryst. Growth \textbf{301-302}, 722-725 (2007)
	\bibitem{Groiss1} H. Groiss, I. Daruka, K. Koike, M. Yano, G. Hesser, G. Springholz, N. Zakharov, P. Werner, F. Sch{\"a}ffler, APL Materials \textbf{2}, 012105 (2014)
	\bibitem{Karczewski1} G. Karczewski, M. Szot, L. Kowalczyk, S. Chusnutdinow, T. Wojtowicz, S. Schreyeck, K. Brunner, C. Schumacher, L. W. Molenkamp, Nanotechnology \textbf{26}, 135601 (2015)
	\bibitem{Minkowski1} M. Mi{\'n}kowski, M. A. Za{\l}uska-Kotur, {\L}. A. Turski, G. Karczewski, J. Appl. Phys. \textbf{120}, 124305 (2016)
	\bibitem{Minkowski2} M. Mi{\'n}kowski, M. A. Za{\l}uska-Kotur, S. Kret, S. Chusnutdinow, S. Schreyeck, K. Brunner, L. W. Molenkamp, G. Karczewski, J. Alloys and Compounds \textbf{747}, 809-814 (2018)
	\bibitem{Bukala1} M. Buka{\l}a, P. Sankowski, R. Buczko, P. Kacman, Nanoscale Research Letters \textbf{6}, 126 (2011)
	\bibitem{Leitsmann1} R. Leitsmann, L. E. Ramos, F. Bechstedt, Phys. Rev. B \textbf{74}, 085309 (2006)
	\bibitem{Chonan1} T. Chonan, S. Katayama, J. Phys. Soc. Jpn. \textbf{75}, 064601 (2006)
	\bibitem{Qiu1} B. Qiu, H. Bao, X. Ruan, ASME 2008 3rd Energy Nanotechnology International Conference, 2008, p. 45
	\bibitem{Wang1} Z.Q. Wang, D. Stroud, A.J. Markworth, Phys. Rev. B \textbf{40},  3129 (1989)
	\bibitem{Oh1} J. Oh, C.H. Grein, J. Cryst. Growth \textbf{193}, 241 (1998)
	\bibitem{Behler1} J. Behler, M. Parrinello, Phys. Rev. Lett. \textbf{98}, 146401 (2007)
	\bibitem{Behler2} J. Behler, J. Phys.: Condens. Matter \textbf{26}, 183001 (2014)
	\bibitem{Behler3} J. Behler, J. Chem. Phys. \textbf{145}, 170901 (2016)
	\bibitem{Behler4} J. Behler, R. Marto{\v{n}}{\'a}k, D. Donadio, M. Parrinello, Phys. Rev. Lett. \textbf{100}, 185501 (2008)
	\bibitem{Vasp} G. Kresse and J. Furthm{\"u}ller,  Phys. Rev. B \textbf{54}, 11169 (1996) 
	\bibitem{Perdew} J. P. Perdew, A. Ruzsinszky, G. I. Csonka, O. A. Vydrov, G. E. Scuseria, L. A. Constantin, X. Zhou, and K. Burke, Phys. Rev. Lett. \textbf{100}, 136406 (2008)
	\bibitem{Khaliullin1} R. Z. Khaliullin, H. Eshet, T. D. K{\"u}hne, J. Behler, M. Parrinello, Phys. Rev. B \textbf{81}, 100103(R) (2010)
	\bibitem{Morawietz1} T. Morawietz, A. Singraber, C. Dellago, J. Behler, Proc. Natl. Acad. Sci. U. S. A. \textbf{113}, 8368-8373 (2016)
	\bibitem{Singraber1} A. Singraber, T. Morawietz, J. Behler, C. Dellago, J. Phys.: Condens. Matter \textbf{30}, 254005 (2018)	
	\bibitem{Bali1} A. Bali, R. Chetty, A. Sharma, G. Rogl, P. Heinrich, S. Suwas, D. K. Misra, P. Rogl, E. Bauer, R. C. Mallik, J. Appl. Phys. \textbf{120}, 175101 (2016)
	\bibitem{Strauss1} A. J. Strauss, Rev. Phys. Appl. (Paris) \textbf{12}, 167-184 (1977)
	\bibitem{Leitsmann2} R. Leitsmann, F. Bechstedt, Semicond. Sci. Technol. \textbf{26}, 014005 (2011)
	\bibitem{Li1} W.-F. Li, C.-M. Fang, M. Dijkstra, M. A. van Huis, J. Phys.: Condens. Matter \textbf{27}, 355801 (2015)
	\bibitem{Roehl1} J. L. Roehl, S. V. Khare, Solar Energy \textbf{101}, 245-253 (2014)
	\bibitem{Singraber2} A. Singraber, J. Behler, C. Dellago, J. Chem. Theory Comput. \textbf{15}, 1827-1840 (2019)
	\bibitem{Plimpton1} S. Plimpton, J. Comput. Phys. \textbf{117}, 1-19 (1995)
	\bibitem{Behler5} J. Behler, J. Chem. Phys. \textbf{134}, 074106 (2011)
	\bibitem{Singraber3} A. Singraber, T. Morawietz, J. Behler, C. Dellago, J. Chem. Theory Comput. \textbf{15}, 3075-3092 (2019)
	\bibitem{Kalman1} R. E. Kalman, J. Basic. Eng. \textbf{82}, 35-45 (1960)
	\bibitem{Kalman2} R. E. Kalman, R. S. Bucy, J. Basic. Eng. \textbf{83}, 95-108 (1961)
	\bibitem{Smith1} G. L. Smith, S. F. Schmidt, L. A. McGee, Technical Reports R-135 (1962)
	\bibitem{Singhal1} S. Singhal, L. Wu in "Advances in Neural Information Processing Systems 1"; Ed. D. S. Touretzky; Morgan-Kaufmann; p. 133-140 (1989)
	\bibitem{Qian1} H. Qian, M. P. Sheetz, E. L. Elson, Biophys. J. \textbf{60}, 910-921 (1991)
	\bibitem{Michalet1} X. Michalet, Phys. Rev. E \textbf{82}, 041914 (2010)
	\bibitem{Newman1} M. E. J. Newman, G. T. Barkema, "Monte Carlo Methods in Statistical Physics", Oxford University Press (1999)
	\bibitem{Rapaport1} M. C. Rapaport, "The Art of Molecular Dynamics Simulation", Cambridge University Press (1995)

	\bibitem{Ma1} J. Ma, J. Yang, S.-H. Wei, J. L. F. Da Silva, Phys. Rev. B \textbf{90}, 155208 (2014)
	\bibitem{Yang1} J.-H. Yang, J.-S. Park, J. Kang, S.-H. Wei, Phys. Rev. B \textbf{91}, 075202 (2015)
	\bibitem{Gomez1} M. P. Gomez, D. A. Stevenson, R. A. Huggins, J. Phys. Chem. Solids \textbf{32}, 335-344 (1971)
	\bibitem{Wolf1} H. Wolf, F. Wagner, Th. Wichert, Phys. Rev. Lett. \textbf{94}, 125901 (2005)
	\bibitem{Jonsson1} H. Jonss{\'o}n, G. Mills, K. W. Jacobsen, in "Classical and Quantum Dynamics Condensed Phase Simulations", edited by B. J. Gerne, G. Ciccotti, D. F. Coker (World Scientific, Singapore, 1998), p. 385
\end{thebibliography}
\end{document}